\documentclass[useAMS,usenatbib,usegraphicx]{mn2e}

\title{Numerical estimation of densities}
\author[Y.~Ascasibar and J.~Binney]
{Y.~Ascasibar$^{1, 2}$\thanks{E-mail: yago@thphys.ox.ac.uk} and J.~Binney$^1$\\
 $^1$Theoretical Physics, 1Keble Road, Oxford OX1 3NP\\
 $^2$Harvard-Smithsonian Center for Astrophysics, 60 Garden St., Cambridge, MA02138, USA
}

\newcommand{\Revised}[1]{#1}

\newcommand{\be}{\begin{equation}}
\newcommand{\ee}{\end{equation}}
\newcommand{\bea}{\begin{eqnarray}\nonumber}
\newcommand{\eea}{\end{eqnarray}}
\newcommand{\F}{{\sc FiEstAS}}

\newcommand{\sph}{\hat\rho_{\rm SPH}}
\newcommand{\x}{\hat\rho/\rho}
\newcommand{\xf}{\hat{f}/f}
\newcommand{\nn}{N_{\rm n}}
\newcommand{\Mk}{M_{\rm k}}
\newcommand{\mmp}{m_{\rm p}}
\newcommand{\msun}{{\rm M}_\odot}
\newcommand{\Mpc}{\,{\rm Mpc}}
\newcommand{\kms}{\,{\rm km\,s}^{-1}}
\newcommand{\vv}[1]{\bmath{#1}}
\newcommand{\dd}{{\rm d}}

\begin{document}

\maketitle

\begin{abstract}
We present a novel technique, dubbed \F, to estimate the underlying
density field from a discrete set of sample points in an arbitrary
multidimensional space.  \F\ assigns a volume to each point by means
of a binary tree.  Density is then computed by integrating over an
adaptive kernel.

As a first test, we construct several Monte Carlo
realizations of a Hernquist profile and recover the particle density
in both real and phase space. At a given point, Poisson noise causes
the \Revised{unsmoothed estimates} to fluctuate by a factor $\sim2$
regardless of the number of particles.
\Revised{This spread can be reduced to about $1\,$dex ($\sim26$ per cent) by
our smoothing procedure.
The density range over which the estimates are
unbiased widens as the particle number increases}.
Our tests show that \Revised{real-space densities obtained} with an SPH kernel are
significantly more biased than those yielded by \F.
\Revised{In phase space, about ten times more particles are required in order
to achieve a similar accuracy}.

As a second application we have estimated phase-space densities in a
\Revised{dark matter} halo from a cosmological simulation. We confirm the results of
\cite{Arad04} that the highest values of $f$ are all associated with
substructure rather than the main halo, and that the volume function
$v(f)\sim f^{-2.5}$ over about four orders of magnitude in $f$. We
show that a modified version of the toy model proposed by Arad et al.\
explains this result and suggests that the departures of $v(f)$ from
power-law form are not mere numerical artefacts.  We conclude that our
algorithm accurately measure the phase-space density up to the limit
where discreteness effects render the simulation itself unreliable.
Computationally, \F\ is orders of magnitude faster than the method
based on Delaunay tessellation that Arad et al.\ employed, making it
practicable to recover smoothed density estimates for \Revised{sets of
$10^9$ points in 6 dimensions}.
\end{abstract}


  \section{Introduction} \label{secIntro}

The estimation of a continuous density field from a set of discrete
sample points is a common problem.  For example, the estimation of the
matter density in real space is fundamental in numerical studies of
cosmic structure formation, and lies at the heart of N-body codes that
solve the Poisson equation, such as ART \citep{ART97} or MLAPM
\citep{MLAPM01}. It also plays an important r\^ole in hydrodynamic
codes based on Smooth Particle Hydrodynamics (SPH) \citep[see
e.g.][and references therein]{Monaghan92,Gadget02}.

Several problems of current interest involve the estimation of the
density of points in phase space
\citep[e.g.,][]{Binney04,Arad04}.
Although violent relaxation and phase-mixing tend to disrupt
substructures, both stars and dark matter may retain information about
their initial conditions for a relatively long time.  In principle, it
would be possible to trace the merging history of our galaxy by
looking for phase-space streams in the stellar population
\citep[see e.g.][]{HelmiWhite99}.
These streams might be identified as regions of enhanced density in
either real space, phase space, or integral space
\citep{Helmi03}.

Most algorithms for estimating densities require the imposition of a metric
on the space occupied by sample points -- that is, they assume that the
`distance' between every two points is defined.
\Revised{However, often the underlying physical problem does not have any metric.
In such a case, a metric can be imposed by specifying a matrix $K_{ij}$ that
relates one unit along the $i$-th axis to one unit on the $j$-th axis.
For example, in the case of phase space, we need a dimensional constant $K$ to
relate positions and velocities, so that a velocity difference $v$ is equivalent
to a spatial offset of $Kv$.
The local velocity dispersion $\sigma$ would thus define a length scale $K\sigma$
that will in general bear no relation to the actual length scale $l$.
For example, in a singular isothermal sphere, $l\propto r$ while $K\sigma$ is constant.
This effect will be present in many applications, amongst them the
estimation of phase-space densities within dark matter haloes.

In this paper we present} an
algorithm for density estimation that does not require a metric.
We show that \Revised{our method} provides fast and accurate estimates of
both real-space and phase-space densities.
The algorithm's independence of the existence of a metric makes it
applicable to a wide class of problems in statistics \citep[see e.g.][for a review]{Silverman86}, in which one
wishes to find clusters of data points in a space in which
different axes represent quantities with different physical
dimensions.
For example, a data point could be composed of a position, a redshift
and luminosities in several bands, and one wishes to find groupings of
physically related points.
\Revised{Some automated classification algorithms based on probability
densities are currently being applied to large galaxy surveys 
\citep[e.g.][]{Richards_04} that could tremendously benefit from the possibility of working in a completely arbitrary space}.

When a metric is available for the sample space, estimators based on
kernels are widely used. For example, in SPH
\citep{Lucy77,GingoldMonaghan77} one obtains the density from
 \be
\sph(\vv{r})=\sum_{p=1}^N \mmp W(\vv{r}-\vv{r_p},h),
\label{eqSPH}
\ee
where $\mmp$ is the mass of each particle, $\vv{r_p}$ its position, and
$W(\vv{r},h)$ is a kernel characterized by a {smoothing scale}, $h$.
Adaptive resolution is achieved by setting $h$ to the distance to the
$\nn$-th nearest neighbour \citep{HernquistKatz89}.

A promising alternative to equation (\ref{eqSPH}),
\Revised{Voronoi tessellations \cite[see e.g.][]{Okabe92} have been used in
astrophysics to identify overdensities in a Poissonian distribution of
X-ray photons \citep{EbelingWiedenmann93}, as well as detecting galaxy
clusters in observational surveys
\citep[e.g.][]{Ramella01,Kim02,Marinoni02}.  Similar in spirit, the
Delaunay Tessellation Field Estimator \citep[DTFE,][and references
therein]{Pelupessy03} estimates the densities of a set of points from
the volume of the Delaunay cells they belong to. A continuous field can be
obtained by linear interpolation \citep{BernardeauWeygaert96}}.
\citet{Arad04} have recently applied the DTFE method to the
estimation of six-dimensional phase-space densities in N-body
experiments.  They find that the volume distribution function, $v(f)$,
for relaxed haloes follows a power law over
more than 4 decades in $f$.  For high values of the phase-space
density, $v(f)$ is dominated by `cold' substructure rather than by the
parent halo.

\Revised{However, the definition of distance underlying a Voronoi or
Delaunay tessellation is not at all obvious in phase space.
The chosen metric determines which particles are close neighbours, and
thus the resulting tessellation and the densities that are assigned to each particle.
The impact of such an ad-hoc choice is extremely difficult to quantify,
and it would require a problem-by-problem study.
It is easy to understand, though, that one's choice may be of crucial importance in
the general case, in which very different scales (e.g. a halo with substructure)
or obviously non-Euclidean spaces (e.g. number density of star-forming galaxies per stereoradian per redshift per morphological type per luminosity) are invloved.
On the other hand, we would like the results to be as stable as possible if we decided to work in different units (e.g. comoving distance instead of redshift, or absolute magnitude instead of luminosity).
No linear scaling $K_{ij}$ could accommodate such a transformation.
The choice of units and coordinate types (e.g. radial, polar, logarithmic) is critical in metric-based schemes.

We present here a non-metric Field Estimator for Arbitrary Spaces
(\F), thoroughly} described in Section~\ref{secAlgor}.  Tests of the algorithm are
presented in Section~\ref{secH}, where we attempt to recover both the
real- and phase-space density from random realizations of a
\citet{Hernquist90} profile, for which an analytical distribution
function is known.  We investigate the phase-space structure of a
realistic dark matter halo from an N-body simulation in
Section~\ref{secSim}, and discuss the computational performance of our
method in Section~\ref{secPerf}.  Section~\ref{secConclus} summarizes
our main conclusions.

  \section{The Algorithm} \label{secAlgor}


\subsection{Tessellation}

Like the DTFE method, \F\ is based on a {tessellation} of the
$d$-dimensional space, i.e. a division of $R^d$ in into mutually
disjoint polygons.  However, we do not resort to a Delaunay or Voronoi
tessellation.
Although these can adapt very efficiently to the geometry of the problem, they require a metric \Revised{to define distances.
Moreover, they become} computationally expensive for large datasets, particularly when $d$ is large.

Instead, we use a binary tree to tessellate our hyper-volume by
means of a recursive procedure.  Given a set of $N$ points, we split
the space along the $i$-th coordinate.  In the ideal case, \Revised{one} would
classify the points according to their position relative to the median
$x_i$, thus obtaining two partitions with $N/2$ points each.  In
practice, we select the two closest points (one from each side) to the
mean $\left<x_i\right>$ and split the $i$-th axis at $x_{\rm
split}=(x_++x_-)/2$.  This is sufficiently accurate for our purposes,
and much more efficient than computing the median.

The process is repeated for both subsets until the initial space is
divided in $N$ hyper-boxes, each containing a single point.  A first
estimate of the local density is just the weight of the point, which
we shall call its mass $\mmp$, divided by the volume of its hyper-box,
\be
\hat \rho_p= \frac{\mmp}{\hat V_p}.
\label{eqRho}
\ee

\begin{figure}
  \centering \includegraphics[width=8cm]{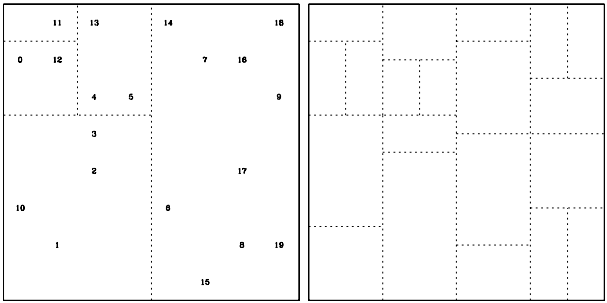}
  \caption{
Construction of a two-dimensional binary tree.
The first division occurs along the horizontal axis, and $x_{\rm split}$ is set between points 5 and 14.
The left part is then divided along the vertical axis (between points 5 and 3), and so on.
  }
  \label{figBT}
\end{figure}

The construction of a two-dimensional binary tree is illustrated in
Fig.~\ref{figBT}.
\Revised{In the general case, the order in which the coordinates are selected for splitting should be chosen at random.
Nevertheless, we try to respect the symmetries of our particular problem (the estimation of phase-space densities) as much as possible, so we alternately split a real-space and a velocity-space coordinate.
In both cases, we always select the axis with the highest elongation, $\left<x_i^2\right>-\left<x_i\right>^2$}.
Proceeding in this way does not significantly affect the estimated densities, but it alleviates numerical problems that arise when two points have very similar (or equal) values of a coordinate $x_i$.
\Revised{It also yields cells that are more or less cubic when projected
onto the position or velocity sub-spaces, which  both have Euclidean metrics.
However, we do not impose any relation between these sub-spaces, nor do we attempt to compute any `distance' involving more than one coordinate at a time.

As discussed above, Voronoi or Delaunay tessellations are not well defined in spaces lacking a metric.
Our scheme is not be sensitive to the particular choice of units, the characteristic scale (if any) or the nature of each dimension, although the precise shape of our orthohedral cells depends on the choice of principal axes (for those sub-spaces admitting rotations) as well as on the splitting sequence.
None the less, the uncertainity due to this non-uniqueness is comparable to or lower than the intrinsic Poisson noise of the point distribution.


\subsection{Boundary effects}

A problem common to all tessellations (with Delaunay triangulation probably yielding the smallest errors) is the choice of a boundary for the system.
Since the density field can be completely arbitrary, there is no reason to expect that the point distribution under study should fit into an orthohedral bounding box.

In particular, there is no warranty that the field will be well sampled near the boundaries.
This is clearly seen, for example, if we take a finite distribution, fully contained within the bounding hyper-box, and enlarge the box without altering the data points themselves.
Most algorithms, including ours, will be stable in the innermost regions, where
every point is surrounded by neighbours (i.e. the field is well
sampled), but densities in the outskirts will be lower
because each point would be associated with a larger volume.
This phenomenon is not an issue when we are interested in a small subset of the point distribution or when periodic boundary conditions apply.
Then, boundary effects are hidden within the Poisson noise.
However, we expect to encounter this difficulty fairly often, since it arises in any distribution sampling an infinite space, specially when there is a sharp density gradient in the outermost regions.
In the estimation of phase-space densities, it manifests most acutely for
particles in the tails of the velocity distribution}.

We try to compensate this effect by redefining the boundary of the
system.  Consider, for example, a sample point that lies at
$\vv{x}^{\rm p}$ within a box that extends to the current boundary of
the sampled region, in a portion of the hyperplane $x_i=X_{\rm max}$.
Let the opposite (interior) face of the point's volume be defined by
the hyperplane $x_i=X_{\rm min}$.  Then we bring the outer face of the
box \Revised{from $X_{\rm max}$ into the hyperplane
$x_i=x_i^{\rm p}+(x_i^{\rm p}-X_{\rm min})$}.
This redefinition of the outer face
of the box reduces its volume and thus increases the derived
density. The changes are small if the field is well sampled near the
boundaries, because then the point will be, on average, in the middle
of the original box.  If the field is severely undersampled, $x^p_i$
will lie much nearer to $X_{\rm min}$ than $X_{\rm max}$ and the
change will be more significant.  The correction is a crude attempt to
adapt the shape of the boundary of the volume within which densities
are determined to the region of space that is usefully sampled.


\subsection{Smoothing}

The prescription outlined above provides a very fast estimation of the
density.  It has the additional advantage that volume integrals of the
type
\be
I=\int_V \Psi[\rho(\vv{x})]~\dd^d\vv{x}
\ee
are readily computed as
\be
\hat I=\sum_{p=1}^N \Psi(\hat\rho_p) \times (\hat V_p \cap V).
\label{eqI}
\ee
 However, the estimator (\ref{eqRho}) is prone to a large statistical error.
Even worse, the associated uncertainty does not vanish, no matter how many
points we use.  As noted by \citet{Arad04}, the probability distribution
$p(\x)$ does not approach a Dirac delta function, even in the limit
$N\to\infty$.  This problem is common to all methods based on a fixed number
of points (in our case, just one).  At constant density, increasing the
total number of points yields smaller and smaller volumes.  Since the
process is scale-invariant, both the probability distribution $p(\x)$ and
the relative error $\Delta\x$ remain constant.

\Revised{Ideally one has just enough particles that the density
changes by only a small amount between neighbouring particles.
Increasing the resolution beyond this ideal does not of itself reduce the error in the
measured density, but it does make it possible to reduce the statistical
error by \emph{smoothing} our density estimate (\ref{eqRho})}.
\F\ implements a kernel interpolation, similar in spirit to the SPH
scheme.  For a given point $\vv{x}$ in $d$-dimensional space, we
integrate the mass over a region $V$ around $\vv{x}$.  First, the mass
of each data point is uniformly distributed within its assigned
volume, $\hat V_p$.  Then, using equation (\ref{eqI}) we calculate \be
\hat M(\vv{x},V)=\sum_{p=1}^N \hat\rho_p \times (\hat V_p \cap V),
\label{eqM}
\ee
and we set
\be
\Revised{\hat \rho(\vv{x},V)=\frac{\hat M(\vv{x},V)}{V}}
\label{eqRho2}
\ee
The shape of $V$ is chosen to be orthohedral, both because of
computational efficiency and because we do not assume the existence
of a metric for the space\footnote{\Revised{It would be hard to define any other shape (e.g. a hyper-sphere) without specifying a metric.}}.We adjust the volume $V$ as described
below such that $\hat M(\vv{x},V)=\Mk$, where the \emph{kernel mass}
$\Mk$ is the only free parameter in \F.  Basing on our preliminary
tests, we suggest $\Mk\simeq10\mmp$ as a compromise between accuracy
and spatial resolution.

\begin{figure}
  \centering \includegraphics[width=6cm]{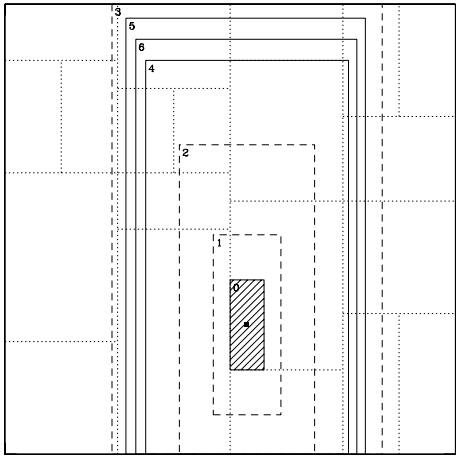}
  \caption{
Smoothing around the position of point 6 in Fig.~\ref{figBT}.
We start with the shadowed box, defined by \Revised{the nearest faces of the cell.
We double the box size (dashed lines) until $\hat M>\Mk$, and then find the exact solution (thick solid line) by repeatedly halving the search interval (thin solid lines).
Numbers at the top left corner of each box illustrate this sequence}.
  }
  \label{figFindL}
\end{figure}

The boundaries of the volume $V$ are the hyperplanes on which the
$i$-th coordinate equals $x_i\pm\Delta x_i$, where the half-width
$\Delta x_i$ is determined iteratively (see Fig.~\ref{figFindL}).  We
start with $\Delta x_i$ equal to the minimum distance along the $i$-th
axis to an edge of the hyper-box containing $\vv{x}$.  With this
prescription, $\hat M\le \mmp$.  We then double all the $\Delta x_i$
until $\hat M>\Mk$, and then find the exact value by repeatedly
halving the search interval.

  \section{Hernquist profile} \label{secH}

In order to test our algorithm, random realizations of a \citet{Hernquist90}
profile have been generated with different numbers of particles, $N$.
We then try to reconstruct the density field both in real and phase space, assessing the accuracy of our estimation, as well as the importance of smoothing and correcting for boundary effects.

The density in real space is given by
\be
\rho(r)=\frac{M}{2\pi a^3}\frac{1}{r/a(1+r/a)^3},
\label{eqHrho}
\ee
\Revised{
which leads to a cumulative mass
\be
M(r)=M \left( \frac{r/a}{1+r/a} \right)^2
\label{eqMh}
\ee
and a gravitational potential
\be
\Phi(r)=-\frac{GM}{a} \frac{1}{1+r/a}.
\ee

The main advantage of the Hernquist profile, which greatly simplifies our analysis, is that the phase-space density can be evaluated analytically as a function of energy,
\bea
f(E) &=& \frac{M/a^3}{4\pi^3(2GM/a)^{3/2}}
\times\\ &\times&
\frac{ 3\sin^{-1}q+q\sqrt{1-q^2}(1-2q^2)(8q^4-8q^2-3) }
     { (1-q^2)^{5/2} },
\label{eqHf}
\eea
where $M$ is the total mass, $a$ is a scale length and
\be
q\equiv\sqrt{-\frac{E}{GM/a}}.
\ee

The \citet{Hernquist90} model has been widely used to model the mass
distribution of dark matter haloes in numerical simulations, and it provides a
good fit to the observed surface brightness of bulges and elliptical galaxies.
Some other profiles have been proposed in the literature \citep[e.g.][]{NFW97,Moore99} that better fit the results of cosmological N-body simulations, but they lack an analytical expression for the phase-space density.

The actual generation of random N-body realisations is fairly straightforward once the mass profile and the phase-space density are known.
The radial coordinate is chosen by generating a random number $x$ between 0 and 1, and then inverting equation (\ref{eqMh}),
\be
\frac{r}{a}=\frac{\sqrt{x}}{1-\sqrt{x}}.
\ee
Then, velocities are assigned from the distribution
\be
p(v)\,\dd v= \frac{4\pi}{\rho(r)}\ f( {v^2}/{2}+\Phi(r) )\ v^2 \dd v
\ee
using von Neumann rejection \citep[see e.g.][]{NR}.
We generate a tentative velocity $v$, uniformly distributed between 0 and $v_{\rm max}=\sqrt{-2\Phi(r)}$, and an auxiliary random number $x$ between 0 and $v_{\rm max}^2f(v_{\rm max})$.
The velocity is accepted only if $x<v^2f(v)$.
Otherwise, two new random numbers are generated for $v$ and $x$ until a value is finally accepted.
Angular coordinates for both positions and velocities are obtained from the variables $0<\phi<2\pi$ and $-1<\cos\theta<1$}.


\subsection{Real space}
\label{secHrho}

\begin{figure}
  \centering \includegraphics[width=8cm]{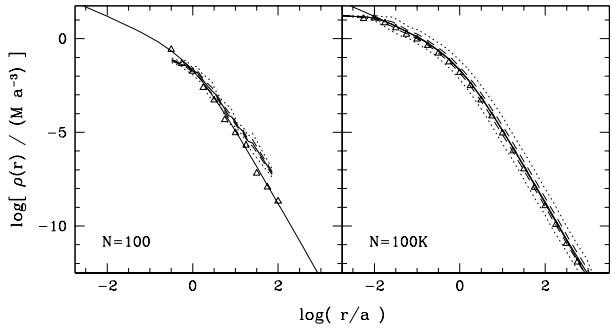}
  \caption{
Reconstruction of a Hernquist density profile from two random realizations with $N=100$ (left) and $N=10^5$ (right) particles.
Thin solid lines show equation~(\ref{eqHrho}).
The average estimation given by \F\ is plotted as a thick solid line.
Dashed and dotted lines represent one and three-sigma scatter, respectively.
\Revised{The density profile derived from particle counts in logarithmic radial bins is shown as open triangles}.
  }
  \label{figHrho}
\end{figure}

The real-space density obtained from two random realizations with
$N=100$ and $N=10^5$ particles is compared in Fig.~\ref{figHrho} with
the theoretical profile (\ref{eqHrho}).
\Revised{We corrected for boundary effects and applied smoothing with $\Mk=10\mmp$ in both cases.
The mean density and standard its deviation were computed in 30 logarithmic bins, subject to the constraint that each bin should contain at least 2 particles}.
For $N=100$, the profile is accurately recovered for \Revised{about} one decade in radius.
In the central regions, lack of resolution erases the density cusp, while in the outskirts of the system, the sampling provided by 100 points is too coarse for a reliable reconstruction of the density field.

\begin{figure}
  \centering \includegraphics[width=7.3cm]{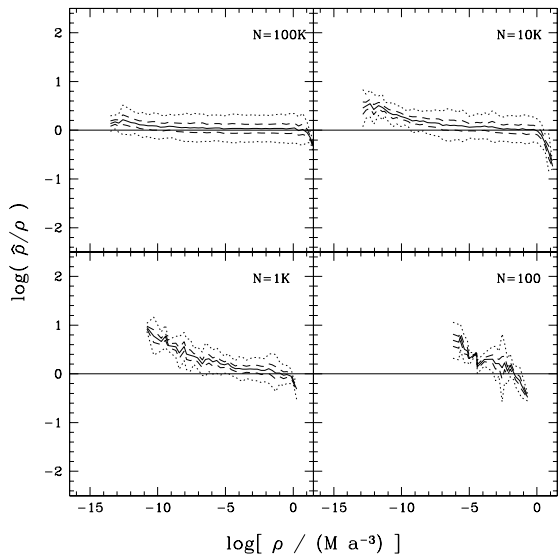}
\caption{
  Systematic dependence of $\x$ on $\rho$ for different numbers of
  particles in a Hernquist sphere.  Boundary correction and smoothing
  have been applied in all cases.  Line styles are the same as in
  Fig.~\ref{figHrho}.}
\label{figHrhoN}
\end{figure}

This can be more clearly seen in Fig.~\ref{figHrhoN}, where we plot
the ratio $\x$ with respect to the theoretical density $\rho$ for
different values of $N$.  In all cases, the estimated density is most
reliable in the intermediate range.  There is some inner radius
$r_{\min}$, set by the number of particles, beyond which we are not
able to resolve the cusp.  On the other hand, there is some outer
radius $r_{\max}$, also set by $N$, where the density becomes so low
that there are simply not enough points to sample the field, and the
estimate $\hat\rho$ is dominated by boundary effects.

\begin{figure}
  \centering \includegraphics[width=7.3cm]{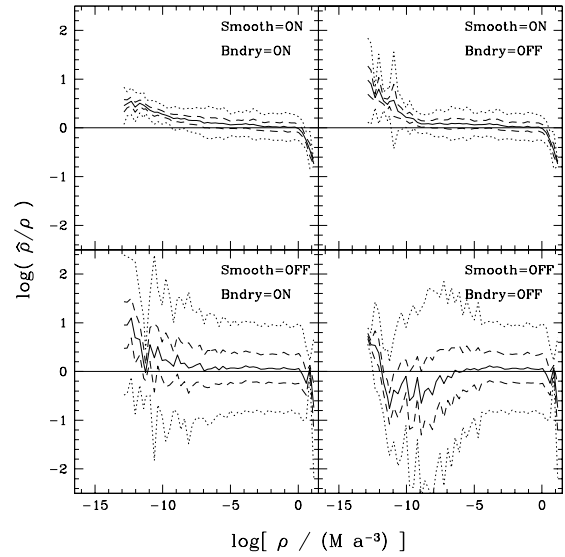}
\caption{
  Systematic dependence of $\x$ on $\rho$ for different prescriptions.
  Density has been smoothed on the top panels.  Left panels apply the
  boundary correction.  All plots correspond to a realization of a
  Hernquist sphere with $N=10^4$ particles.  Line styles are the same
  as in Fig.~\ref{figHrho}.}
\label{figHrhoAlgor}
\end{figure}

Fig.~\ref{figHrhoAlgor} shows that the boundary correction is relevant
for the unsmoothed estimator (bottom panels) at low densities, but it
has a \Revised{weaker} effect on the smoothed estimate (top panels).
For particles at the boundary, the volume assigned by the binary tree is
usually too large, because the bounding box is a cuboid, whereas the
density field has spherical symmetry.  There is a lot of empty space
from the last particle to the appropriate faces of the bounding box,
resulting in an artificially large value of $V_p$.
The density given by expression (\ref{eqRho}) is thus severely
underestimated for these particles.  When the smoothed estimator
(\ref{eqRho2}) is used instead, the effect is more limited, since each
particle will typically be surrounded by higher density neighbours.
These will dominate the volume integral, minimising the contamination
from artificially low-density particles.

\begin{figure}
  \centering \includegraphics[width=7cm]{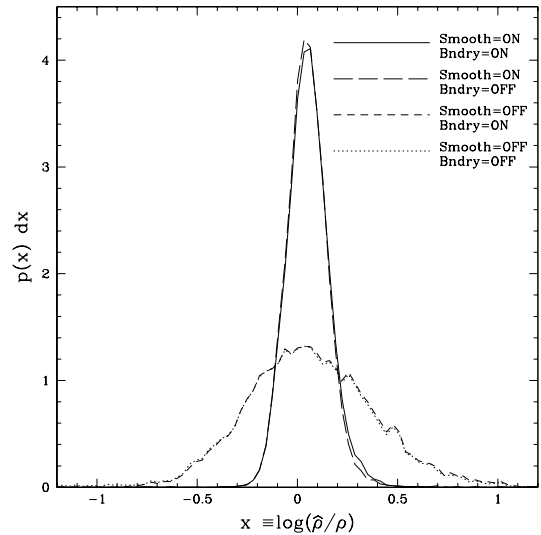} \caption{ Probability
  distribution $p(\x)$ for a Hernquist sphere with $N=10^4$ particles,
  using different prescriptions.  Correction for boundary effects is
  negligible, while smoothing considerably reduces the uncertainty.  }
  \label{figHrhoAlgor2}
\end{figure}

Moreover, the number of particles affected by the boundary correction
constitutes a relatively small fraction of the total.  As can be seen
in Fig.~\ref{figHrhoAlgor2}, the overall distribution of $\x$ is not
altered at all by considering boundary effects, even for moderate
resolutions ($N=10^4$ particles).
This correction becomes more
important for very poor resolutions, or in higher-dimensionality
spaces, in which the fraction of data points at the boundary is
substantially larger.

\begin{figure}
  \centering \includegraphics[width=7cm]{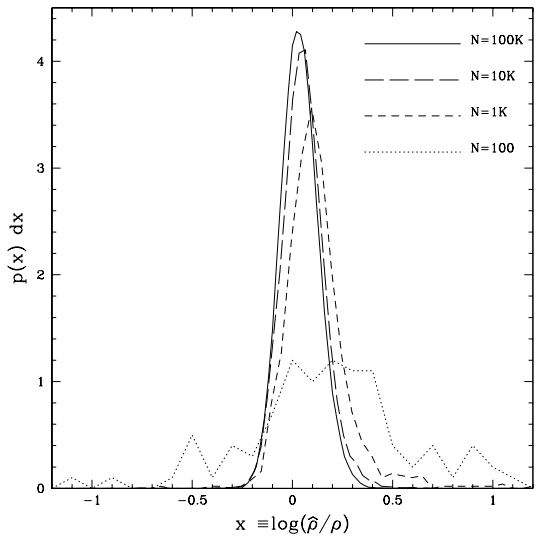} \caption{ Probability
  distribution $p(\x)$ for different numbers of particles in the
  Hernquist sphere.  The distribution is broader and more skewed as
  $N$ decreases.  } \label{figHrhoN2}
\end{figure}

As previously mentioned, the probability distribution $p(\x)$ does not
tend to a Dirac delta function as $N\to\infty$ (see
Fig.~\ref{figHrhoN2}).
\Revised{If our point distribution is locally Poissonian, the dispersion can be reduced by averaging over a neighbouring volume containing more particles.
The underlying density field must be approximately constant (i.e. we must have enough resolution) over the smoothing region}.
This assumption breaks down at the very centre, as well as in the 
outermost parts.
In these regions, smoothing may even worsen the density estimate, but
for most particles, it yields a significant improvement in acuracy
(see Figs~\ref{figHrhoAlgor} and \ref{figHrhoAlgor2}).

It is encouraging that $p(\x)$ varies very little for $N>10^3$
particles, suggesting that it has converged to its asymptotic form.
Our $p(\x)$ is well modelled by a log-normal distribution with
$\left<\log(\x)\right>\le0.04\,$dex and $\sigma\le0.1\,$dex, which
corresponds to a relative error $\Delta\rho/\rho$ smaller than 26 per
cent.

\begin{figure}
\centering \includegraphics[width=7cm]{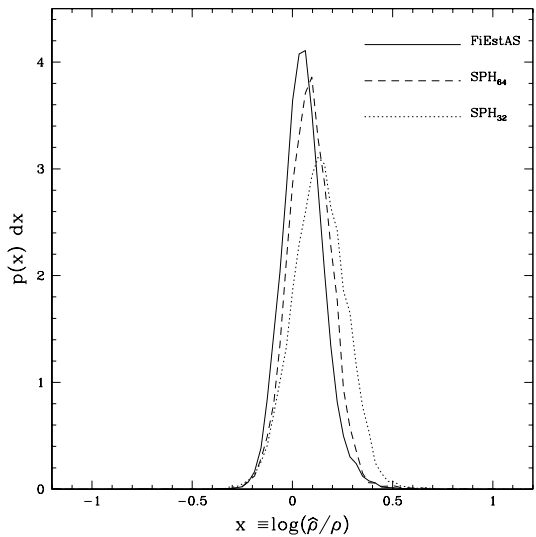} \caption{
  Comparison between \F\ \Revised{(with $Mk=10$)} and SPH (using 32 and 64 neighbours)
  for a Hernquist sphere with $N=10^4$ particles.
}\label{figHSPH}
\end{figure}

In Fig.~\ref{figHSPH} \F\ is compared with the popular SPH method.  We
set the smoothing length in the latter using 32 and 64 neighbours. \F\
gives a slightly more accurate (i.e. lower scatter) estimate, and the
results are significantly less biased towards large $\x$.
\Revised{In order to obtain similar results, the SPH method would require
averaging over more than 64 neighbouring particles, which is considerably
demanding from the computational point of view, and seriously degrades the
spatial resolution of the estimator}.


\subsection{Phase space}

\begin{figure}
  \centering \includegraphics[width=8cm]{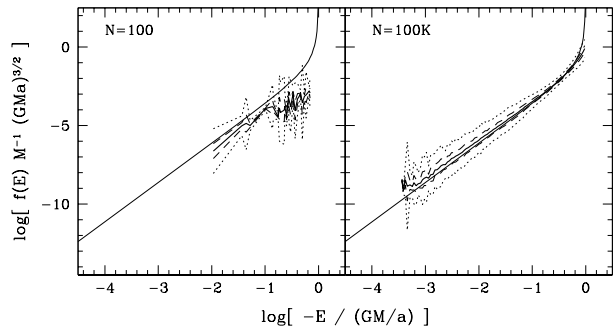} \caption{ Phase-space
  density of two random realizations of a Hernquist model, with
  $N=100$ (left) and $N=10^5$ (right) particles.  Line styles are the
  same as in Fig.~\ref{figHrho}.  Thin solid lines represent
  equation~(\ref{eqHf}).  } \label{figHf}
\end{figure}

In this section, we test the performance of our algorithm in
six-dimensional phase space.
\Revised{From the distribution function (\ref{eqHf}) we
can randomly sample the phase space with any number $N$ of particles.
Fig.~\ref{figHf} shows the results for two realizations of a
Hernquist model with different values of $N$.
For $N=100$, we are barely able to recover the qualitative behaviour
of the theoretical profile, while with $N=10^5$ the points scatter around
the analytic value over eight orders of magnitude in $f$}.

\begin{figure}
  \centering \includegraphics[width=7.3cm]{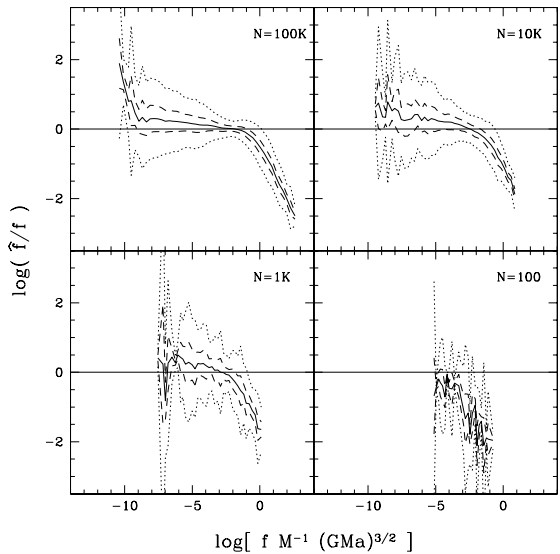}
  \caption{
    Dependence of $\xf$ on $f$ for a Hernquist model sampled with various $N$.
    \Revised{Smoothing and boundary corrections are both on}.
  }\label{figHfN}
\end{figure}

\begin{figure}
  \centering \includegraphics[width=7cm]{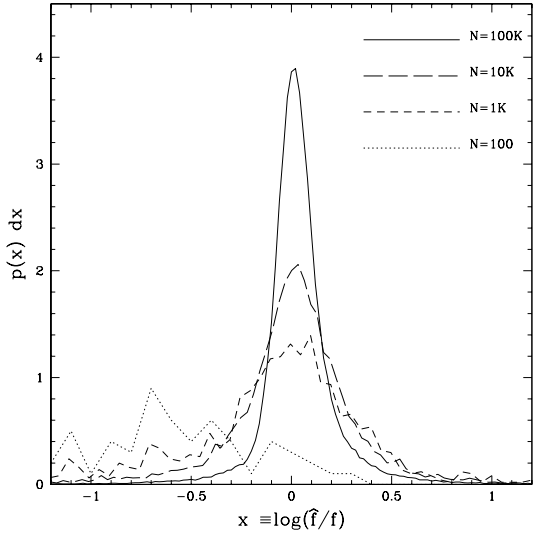} \caption{
  Probability distribution $p(\xf)$ for a Hernquist model sampled with
  various $N$.
  \Revised{Smoothing and boundary corrections are both on}.
} \label{figHfN2}
\end{figure}

The effect of increasing the number of particles can be appreciated in
Fig.~\ref{figHfN}.  More resolution extends the range of phase-space
densities that are accurately estimated.
\Revised{As in the real-space case, lack of
resolution leads to systematic underestimation of the highest densities,
while there is a tendency to overestimate the lowest densities.
Fig.~\ref{figHfN2} shows that with $N=10^4$ the
probability distribution $p(\xf)$ is reasonably unbiased,
$\left<\log(\xf)\right>\le0.04\,$dex, and has a typical spread
$\sigma\sim0.1\,$dex comparable to that in real space.
Note however that more resolution is obviously needed in order to assess convergence.
Comparison with the real-space densities shown in
Figs.~\ref{figHrho}, \ref{figHrhoAlgor} and \ref{figHrhoN2} suggests that
increasing the dimensionality of the space from 3 to 6 requires a factor of
order 10 more points to achieve a similar distribution.
In general, it seems reasonable to expect that the number of points needed to
properly sample a given space increases exponentially with the number of
dimensions $d$ in a way similar to the typical number of neighbours, roughly $2^d$}.

\begin{figure}
  \centering \includegraphics[width=7.3cm]{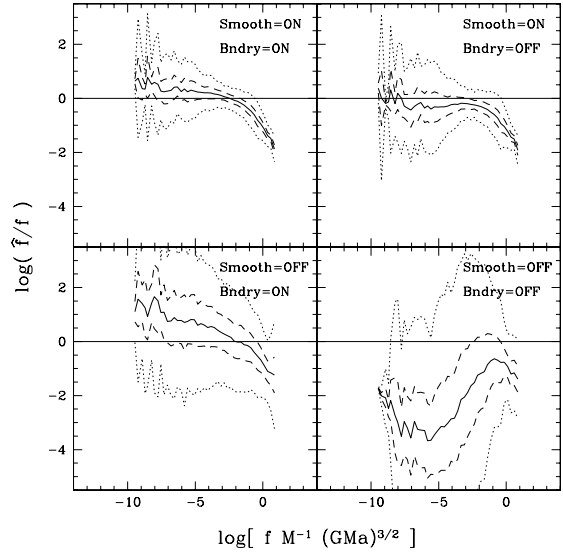} \caption{
  Dependence of $\xf$ on $f$ for a Hernquist model of $N=10^4$
  particles when various combinations of smoothing and boundary
  correction are used.  } \label{figHfAlgor}
\end{figure}

We quantify in Fig.~\ref{figHfAlgor} the importance of smoothing and
boundary corrections.  In six dimensions, a very large fraction of
particles lie close to at least one of the 12 faces of the bounding
hyper-box.  The correction for boundary effects is thus relevant even
for $N=10^4$.  Using the smoothed estimator considerably reduces the
dispersion in $\log(\xf)$, but $\hat{f}$ is still biased low when
boundary effects are not accounted for.  There are so many particles
whose initial density (\ref{eqRho}) was underestimated that they make
a significant contribution to the volume integral in (\ref{eqRho2}).

Another effect of the coarser sampling in six dimensions is that the
softening of the cusp is more noticeable than in the three-dimensional
case.
Note also that the slope of the phase-space density for $E\to0$ is much
steeper than the physical density at low $r$, which makes it difficult to
sample the high-$f$ region adequately.

\begin{figure}
  \centering \includegraphics[width=7cm]{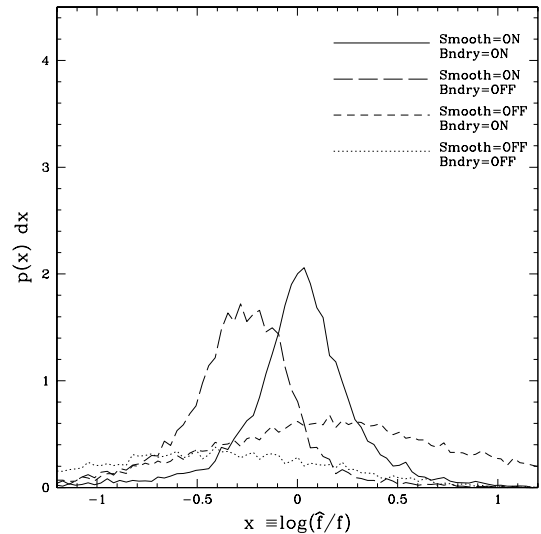} \caption{
  Probability distribution $p(\xf)$ for a Hernquist model of $N=10^4$
  particles when various combinations of smoothing and boundary
  correction are used.  } \label{figHfAlgor2}
\end{figure}

\Revised{None the less, the smoothed estimator (\ref{eqRho2}) greatly
reduces the spread in $\hat{f}/f$}.
As can be seen in Fig.~\ref{figHfAlgor2}, the unsmoothed estimator yields a dispersion
of about one decade around the true value of $f$, regardless of
whether the boundary effects have been corrected or not.
Such a large dispersion coincides approximately with the statistical
error reported by \citet{Arad04}.

Arad et al.\ investigated the volume distribution function \be
v(f_0)=\int \delta[f(\vv{x},\vv{v})-f_0]\ \dd^3\vv{x}\ \dd^3\vv{v},
\ee
 which gives the volume of phase space in which $f$ lies in the
 interval $(f_0,f_0+\dd f)$, i.e. $\dd V=v(f)\,\dd f$.  Since the mass
 contained in particles that have phase space densities within $\dd f$
 of $f$ is \be
\frac{\dd m}{\dd f}=fv(f),
\ee
we can estimate $v(f)$ from $\dd m/\dd f$. We obtain the latter by
sorting the particles in order of increasing $\hat f$-values and then
binning them. In our tests the bins contain 50 particles each, except
in the case $N=100$, when we take only 5 particles per bin.

\citet{Arad04} show that errors in $\hat f$ cause
the estimated volume function $\hat{v}(f)$ to be the convolution \be
\hat{v}(f_0)=\int_0^\infty v(f) p(f_0/f) f^{-1} \dd f.
\ee
For $v(f)\propto f^{-\alpha}$, we find that $\hat{v}(f_0)\propto
f_0^{-\alpha}$, where the constant of proportionality is an integral
that is independent of $f_0$. From this result it follows that when
$v(f)$ is well approximated locally by a power law over the range in
which $p(f/f_0)$ is non-negligible, the measured logarithmic slope of
$\hat{v}(f)$ at $f_0$ will equal the logarithmic slope of the
underlying function $v(f)$.

When $f$ is a function $f(E)$ of energy alone, we obtain the
theoretical volume distribution from
\be
v[f(E)]=\frac{g(E)}{f^\prime(E)},
\label{eqHvf}
\ee
 where $g(E)$ is the density of states and $f^\prime(E)$ denotes the
 derivative of $f$ with respect to $E$.  For a Hernquist profile, $f$
 is given by equation (\ref{eqHf}), and the density of states is \bea
 g(E) &=& \frac{2\pi^2a^3(2GM/a)^{1/2}}{3q^5} [\
 3(8q^4-4q^2+1)\cos^{-1}q
\\ && - 
q(1-q^2)^{1/2}(4q^2-1)(2q^2+3)\ ].
\eea

\begin{figure}
  \centering \includegraphics[width=8cm]{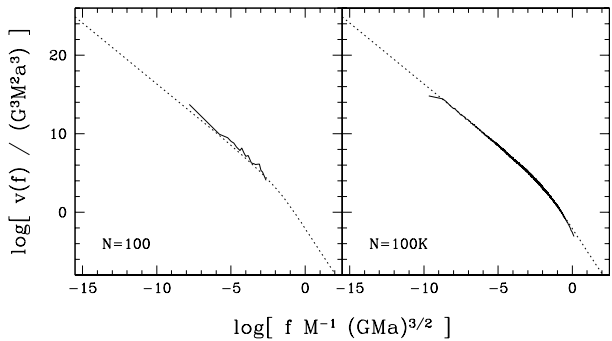} \caption{ Volume
  distribution $v(\hat{f})$ for Hernquist models with $N=100$ and
  $N=10^5$.  The analytic result (\ref{eqHvf}) is plotted as a dotted
  line.  } \label{figHvf}
\end{figure}

\begin{figure}
  \centering \includegraphics[width=7cm]{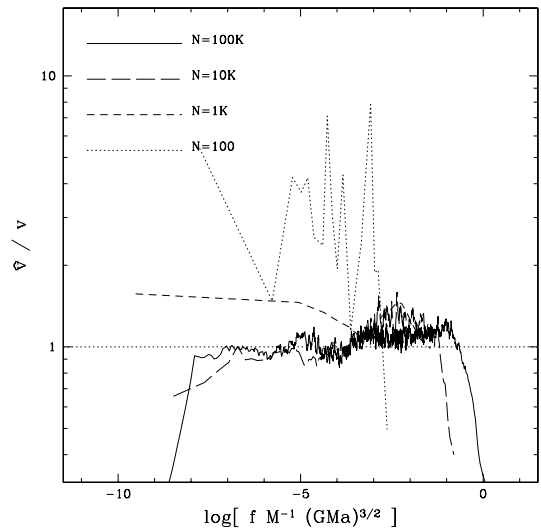} \caption{ Systematic
  dependence of $\hat{v}/v$ on $f$ for Hernquist models with various
  $N$.  } \label{figHvfN}
\end{figure}

Fig.~\ref{figHvf} demonstrates that it is possible to recover the
qualitative behaviour of $v(f)$ even with a resolution as poor as
$N=100$.  Deviations from the theoretical volume distribution are
plotted in Fig.~\ref{figHvfN} for different numbers of particles.
Results for $N\ge1000$ stay within a factor of 2 of the true $v(f)$,
although some systematic errors are present near the turnover of
$f(E)$, where the distribution function cannot be described as a power
law.

  \section{Cosmological simulation} \label{secSim}

Having assessed the accuracy of our algorithm, we apply it to a
realistic case in which the phase-space structure is not known a
priori.  We select a $\sim10^{14}\msun$ halo from an N-body simulation
accomplished with the publicly-available code {\sc Gadget}
\citep{Gadget01,Gadget02}.  The object contains $N=1\,190\,016$ dark
matter particles, embedded in a $80\ h^{-1}$ Mpc box in a $\Lambda$CDM
universe ($\Omega_{\rm dm}=0.3$, $\Omega_\Lambda=0.7$, $h=0.7$).  A
thorough analysis of these numerical experiments can be found in
\citet{Ascasibar03,Ascasibar04sc}.

\begin{figure}
  \centering \includegraphics[width=8cm]{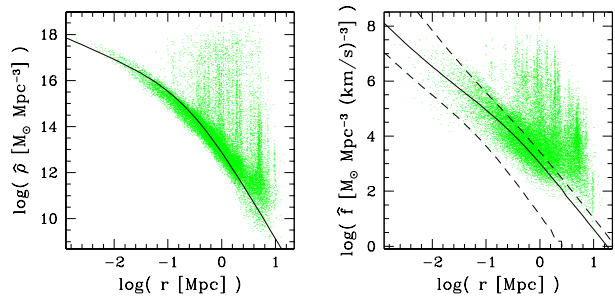} \caption{ Density
  profile, both in real (left) and phase (right) space, of a
  $10^{14}\msun$ halo in a numerical simulation ($N\sim10^6$).  Solid
  lines show a Hernquist model with $M=3.6\times10^{14}\msun$ and
  $a=250\,$kpc.  In phase space, $v^2=3\sigma^2_r$ has been
  assumed. Dashed lines show $v^2=0$ and $v^2=9\sigma^2_r$.  }
  \label{figSrhof}
\end{figure}

A scatter plot of the real- and phase-space density of 2 per cent of
the particles is shown in Fig.~\ref{figSrhof}.  Substructure gives
rise to localized density peaks in both spaces, whereas most of the
mass follows the global trend of decreasing density with radius.  This
component can be \Revised{roughly} described by a Hernquist model with
$M=3.6\times10^{14}\msun$ and $a=250\,$kpc, plotted as solid lines in
the figure.

\begin{figure}
  \centering \includegraphics[width=7cm]{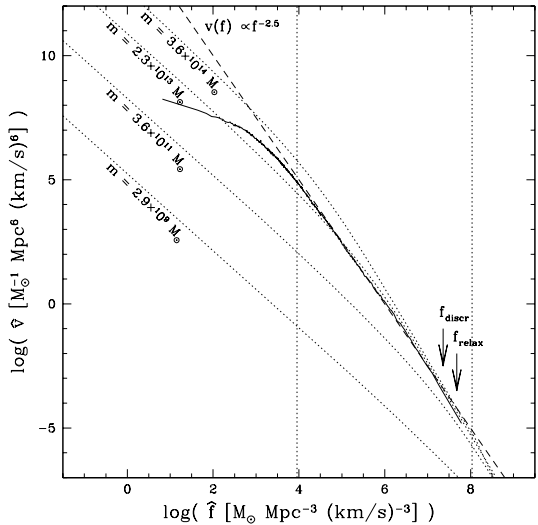} \caption{ Volume
  distribution of phase-space density for a simulated halo (solid
  line).  Dotted lines plot Hernquist models of different masses,
  while the dashed line shows $v(f)\propto f^{-2.5}$.  Vertical lines
  mark the expected turnovers in $v(f)$, and arrows the resolution
  limits of this simulation (see text).  } \label{figSvf}
\end{figure}

Satellite haloes achieve higher central densities than the main
object, and the effect is more pronounced in phase-space.
Therefore, the high end of the volume distribution $v(f)$ will be dominated by substructure.
We plot this quantity in Fig.~\ref{figSvf}, together with the power-law behaviour advocated by \citet{Arad04}.
Although our results are consistent with $v(f)\propto f^{-2.5}$ over 4 decades in $f$, we find departures from this law both at low and high phase-space densities.
\Revised{Similar results are reported by \citet{Arad04}, but deviations from a pure power law are attributed to a lack of resolution at high $f$ and incomplete virialization at low $f$.
We show below that such deviations are indeed expected theoretically for systems in virial equilibrium}.

Also shown in Fig.~\ref{figSvf} are the volume distributions of
several Hernquist models with different masses.
We use the same `concentration' $c\equiv r_{200}/r_{\rm s}=8$ for all
of them, where $r_{200}$ is the radius at which the enclosed density
is 200 times the critical density and $r_s=a/2$ is the point at which
the logarithmic slope of the density profile is equal to $-2$.
With this prescription, the mass and scale length are related by
\be
m = \frac{4\pi}{3}\,200\rho_{\rm c}\,
\frac{c}{2}\left(\frac{c}{2}+1\right)^2 a^3.
\ee

Note that the volume distribution of the Hernquist model tends
asymptotically to $v(f\!\to\!0)\propto f^{-1.56}$
and $v(f\!\to\!\infty)\propto f^{-2.8}$.
The turning point occurs roughly at
 \be
f\sim 3.25\times10^{18}\left(m\over\!\msun\right)^{-1}\msun\Mpc^{-3}(\kms)^{-3}
\ee
and
\be
v\sim 5.46\times10^{-38}\left(m\over\!\msun\right)^3\msun^{-1}\Mpc^{6}(\kms)^{6},
\ee
where $m$ is the mass of the halo.

In the toy model proposed by \citet{Arad04}, the total volume
distribution is given by the convolution
\be
v(f)=\int_0^\infty\!\!\!v_m(f)\,\frac{\dd n(m)}{\dd m}\ \dd m
\label{eqVth}
\ee
of the individual $v_m(f)$ of each sub-halo with the sub-halo mass function
${\dd n}/{\dd m} \propto m^{-1.9}$ \citep{Lucia04}.
It is important, though, that the mass $m$ of the satellites cannot reach
an arbitrary value.
On one hand, the resolution of the simulation imposes a minimum mass
$m_{\rm min}\sim 100 m_{\rm p}$.
On the other hand, there is the physical constraint that $m$ must be
smaller than the mass of the main object.

Let us define
\be
\phi\equiv {f\over3.25\times10^{18}\msun^2\Mpc^{-3}(\kms)^{-3}}
\ee
and approximate $v_m(f)$ as a double power law
\be
v_m(f)\simeq\cases{
 5.46\times10^{-38}m^3 (m\phi)^{-1.56} & for $\phi\le1/m$\cr
 5.46\times10^{-38}m^3 (m\phi)^{-2.8}  & for $\phi\ge1/m$.}
\ee
We can evaluate expression (\ref{eqVth}) as
\bea
v(f) &\propto&
  \phi^{-1.56} \int_{m_{\rm min}}^{1/\phi}\!\!\!m^{-0.46}\dd m
  + \phi^{-2.8}\int_{1/\phi}^{m_{\rm max}}\!\!\!m^{-1.7}\dd m
\\ &=&
  3.28\,\phi^{-2.1}
  - \frac{m_{\rm min}^{0.54}}{0.54}\, \phi^{-1.56}
  - \frac{m_{\rm max}^{-0.7}}{0.7}\, \phi^{-2.8}
\label{eqVth2}
\eea
for $m_{\rm min}<1/\phi<m_{\rm max}$, and
\be
v(f) \propto \phi^\lambda
   \int_{m_{\rm min}}^{m_{\rm max}}\!\!\!\dd m\, m^{1.1+\lambda}
\ee
with $\lambda\!=\!-1.56$ for $\phi\!<\!1/m_{\rm max}$ and
$\lambda\!=\!-2.8$ for $\phi\!>\!1/m_{\rm min}$.

According to this model, the logarithmic slope of the total volume
distribution would vary smoothly from $-1.56$ to $-2.8$ as $\phi$ goes
from $1/m_{\rm max}$ to $1/m_{\rm min}$, the shape of
$v(f)$ being controlled by the values of these parameters.
A constant power law $v(f)\propto f^{-2.1}$ can only be obtained in
the limit $m_{\rm min}\to0$ and $m_{\rm max}\to\infty$.

The phase-space densities implied by
$m_{\rm min}=3\times10^{10}\msun$ and
$m_{\rm max}=3.6\times10^{14}\msun$
are illustrated as vertical dotted lines in Fig.~\ref{figSvf}.
We find strong evidence of the turnover of $v(f)$ at the low end.
The logarithmic slope measured by \F\ is flatter than $-2.1$ over
several orders of magnitude, and it is close to $-1.56$ at the
physical scale dictated by the main halo.
At the high end, the effects of $m_{\rm min}$ are (not surprisingly)
very close to our resolution limit.

Dynamical discreteness effects place an upper limit on
the phase-space density at which a given cosmological simulation is
trustworthy.
These effects include those of granularity when the first
non-linear structures form \citep{Binney04} and two-body relaxation
\citep{Diemand04}.

The phase-space density above which  the two-body relaxation time is
shorter than the age of the universe is
\be
f_{\rm relax} \simeq
\frac{0.34}{(2\pi)^{3/2}G^2\ln\Lambda} \frac{1}{m_{\rm p}t_0}.
\ee
For the simulation studied here, $m_{\rm p}=3\times10^8\msun$ and 
$t_0=13\,$Gyr. Substituting $\ln\Lambda\sim6$,
\be
f_{\rm relax} = 4.8\times10^7\msun\Mpc^{-3}(\kms)^{-3}.
\ee

The phase-space density at which the effects discussed
by Binney become important is 
 \be
f_{\rm discr}\sim \frac{\eta\,m_{\rm p}}{{(H_0\Delta q^2)}^3}
\le \frac{(\Omega_{\rm m}\rho_{\rm c})^2}{H_0^3m_{\rm p}}
\ee
where $\Delta q$ is the
comoving interparticle separation and $\eta$ is a factor a little
smaller than unity that depends on the precise time
at which the first structures become non-linear.
Therefore,
\be
f_{\rm discr}\sim 2.3\times10^7\msun\Mpc^{-3}(\kms)^{-3}.
\ee

Both estimates yield similar values of the maximum phase-space
density that can be reliably represented in an N-body simulation
\citep[for a detailed discussion, the reader is referred
to][]{Binney04}.
The steepening of the volume distribution at the resolution scale is a 
numerical artefact of the simulation, not of our algorithm.
The fact that it is detected and accurately measured suggests that
smoothing and sampling errors in \F\ are significant only at the very
large values of $f$ at which the underlying simulation is dominated by
discreteness effects.

The situation at the low end is far more complicated.
Although the simple model we have considered in this section offers an
interesting insight on the problem, it does not yield a reliable
prediction of the total volume distribution.
The phase-space volumes occupied by the satellites may intersect
amongst themselves, and they obviously do with the phase-space volume
occupied by the main object.
Therefore, the total $v(f)$ of the system is much less
than the sum over the individual sub-haloes, as  equation
(\ref{eqVth}) assumes.
The phase-space density $f_m(\vv{r},\vv{v})$ is additive, but the
volume distribution $v_m(f)$ is not.

Deriving the shape and normalisation of $v(f)$ would require
knowledge of the exact distribution of haloes in phase space.
Qualitatively, we expect that most satellites are located
in the low-$f$ regions of the main halo.
The phase-space density will increase in these regions, most likely
flattening $v(f)$ at the low end.
Hence, logarithmic slopes even flatter than $-1.56$ would not be
unexpected on physical grounds, although there are also many numerical
effects that contribute in the same direction.
Some of them arise from the N-body technique (e.g. the limited number of
high-resolution particles) and some of them are associated with \F\
(e.g. systematic bias at low $f$).
It is difficult to disentangle these numerical artefacts from the genuine
effects of substructure.
Although we cannot establish a definite value for the asymptotic behaviour
of $v(f)$ as $f\to0$,  we do conclude that it is probably flatter than a
substructureless Hernquist sphere, $v(f\to0)\propto f^{-1.56}$.

  \section{Performance}
  \label{secPerf}

State-of-the-art N-body simulations currently yield halos with up to
$N\sim10^7$ particles.
Recent observational archives achieve a similar number of objects.
However, it seems likely that the typical number of points in
future astrophysical datasets will increase by orders of magnitude
during the coming years \citep[see e.g.][]{SzalayGray01}.
Therefore, computational efficiency has become a crucial issue
in data analysis.
A very accurate algorithm is completely useless if it cannot be run in 
a reasonable amount of time.

\begin{figure}
  \centering \includegraphics[width=7cm]{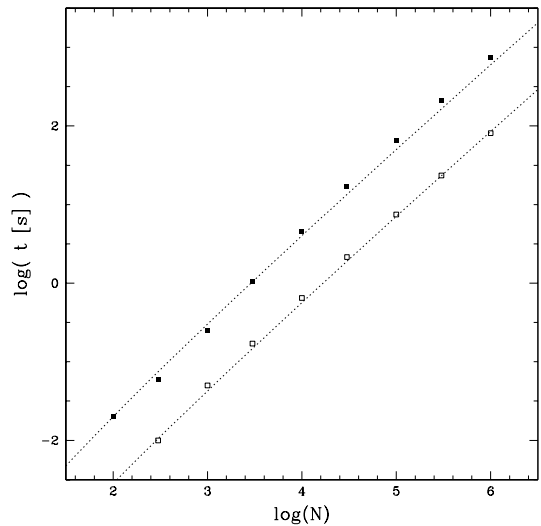}
  \caption{
Computing time versus particle number for the recovery of real-space density
(open symbols) and phase-space density (filled symbols).
Dotted lines mark $N\log(N)$ scaling.
}
  \label{figPerf}
\end{figure}

Fig.~\ref{figPerf} shows, as a function of the number of points $N$, the time
required to recover densities on a garden-variety \Revised{laptop PC (1.7 GHz, 1 Gb RAM)}.
The open symbols show the time required to estimate the
real-space density, while the filled symbols are for recovery of the
phase-space density.
Our algorithm scales as $N\log(N)$, shown as dotted lines, for
$N\ge10^3$ particles.
The offset between real and phase space estimators
arises because the typical number of neighbours a particle has rises with
dimensionality $d$ roughly like \Revised{$2^d$}.

It \Revised{took \F\ $730\,$s to recover $f$ from a set of} $N=10^6$ particles,
whereas \citet{Arad04}
report that about a week is required with a Delaunay tessellation.
In fact the contrast in speed is even more dramatic than this comparison would
suggest, since nearly all the measured time for \F\ is given to smoothing: for
$N=10^6$ the creation of the binary tree requires only $\sim4\,$s
independent of the number of dimensions. The density estimates of Arad et
al.\ are unsmoothed ones and show the same large statistical fluctuations
as our expression (\ref{eqRho}). Hence the week of computational time
reported by Arad et al.\ ought strictly to be compared to the $4\,$s it
takes \F\ to produce unsmoothed density estimates.

  \section{Conclusions}
  \label{secConclus}

In almost every branch of the physical, biological and social sciences one
frequently wishes to determine the density of data points in some space.
As all points in a given cluster are expected to have a common origin,
identifying clusters through maxima in the density of points is likely to
yield clues as to what processes are responsible for distributing points in
the space.
In the general case, each axis will represent a quantity with its
own particular dimensions, and there is no natural definition of the distance
between two data points.

The Field Estimator for Arbitrary Spaces (\F)
that has been presented here provides a fast and accurate way of determining
the density field underlying any given distribution of points.
\F\ obtains a first estimate of the density by associating a volume with
each data point and taking the density to be the inverse of that
volume.
Such an estimate inevitably fluctuates by of order
$1\,$dex regardless of the number of data points, so it is desirable to
smooth it.
The computational cost of \F\ increases with the number of data points $N$ as
$N\log(N)$, and approximately as \Revised{$2^d$ (probably $d\times2^d$)} with
dimensionality $d$.
For $d\la6$ and $N\la10^7$, \Revised{smoothed densities within $\sim0.1\,$dex
error bars (as long as the field is well sampled) are easily recovered
on a single-processor computer}.

The higher the dimension of the space within which densities are required,
the greater is the fraction of the particles that lie near the boundary of
the sampled volume. The determination of the density at the locations of
these particles is problematical \Revised{for any method} because it depends on what
one takes to be the boundary of the sampled volume. We have experimented
with various schemes for defining this boundary, without arriving at a
definitive solution of the problem.

In tests with $d=3$, we find that \F\ returns almost
unbiased densities with $10^5$ points in a Hernquist sphere.
Statistical errors are
$\sim0.1\,$dex over ten orders of magnitude in $\rho$.
Smoothing tends to overestimate the density in low-density regions,
while the opposite is true for the central, high-density region.
Increasing the particle number extends the
range of \Revised{densities} over which \F\ makes an unbiased estimate,
but does not alter the magnitude of the statistical fluctuations in
$\hat \rho$. Our tests show that density estimates obtained
with the popular SPH kernel are significantly more strongly biased (towards
large values) than are estimates from \F.

Most popular density estimators are not well suited to the evaluation of
phase-space densities $f$ because they require the distance between any two
points to be defined and there is no natural definition of distance in phase
space. Since \F\ does not involve the concept of distance, it is well suited
to the estimation of $f$. \Revised{A possible drawback of \F\ is its reliance on a
coordinate grid, so density estimates will vary slightly when the grid is
rotated}. We confirm the results obtained by \citet{Arad04} for a dark halo
from a cosmological simulation, but at greatly reduced computational cost:
the CPU time required is reduced from of order a week for raw densities to
less than \Revised{fifteen minutes} for smoothed densities.
This speedup makes feasible the
determination of phase-space densities in simulations with up to $10^9$
particles.

Although we confirm that over three to five decades in $f$ the
specific volume function follows a power law $v(f)\propto f^{-2.5}$,
we find that at both the lowest and the highest phase-space densities,
$v(f)$ falls below this power law, and we believe this phenomenon is
not an artefact of our analysis.
We are able to account for the form of $v(f)$ with a model in
which a massive CDM halo is a superposition of a population of Hernquist
profiles. \Revised{A similar} model was originally proposed by Arad et al., but in their
analysis it did not reproduce $v(f)$ because they failed to recognize
that in any given simulation there is both a smallest and a largest halo
that can form.

The interest of \F\ extends far beyond the mere estimation of densities;
it opens up a world of possibilities for both
supervised and unsupervised classification.
As an example of the latter, we are currently developing a halo finder
in phase space.
It has the advantage over conventional, real-space based methods that sub-haloes and
tidal streams are very clearly defined in phase space that are
not apparent in real space.
Another possibility would be the identification of galaxy groups and
clusters from observational data.
Not being based on a particular metric, we expect \F\ to provide a
robust parameter-free method that could exploit all the available
information, such as coordinates on the sky, redshifts, colours, etc.
Indeed, \F\ constitutes a general-purpose data-mining
tool with many potential applications, not necessarily restricted to
the field of Astrophysics.


\section*{Acknowledgments}

We thank I. Arad for a useful discussion and C. Nipoti for his
comments on the manuscript.
YA acknowledges support from the \emph{Leverhume Trust} (United Kingdom).

 \bibliographystyle{mn2e}
 \bibliography{../../LaTeX/BibTeX/DATABASE,../../LaTeX/BibTeX/PREPRINTS}

\end{document}